\documentclass[iop]{emulateapj}
\usepackage{apjfonts}

\begin{document}

\shortauthors{Berdyugina et al.}
\shorttitle{Polarized reflected light from HD189733b}

\title{Polarized reflected light from the exoplanet HD189733b:\\ 
First multicolor observations and confirmation of detection}

\author{S. V. Berdyugina\altaffilmark{1}, 
A. V. Berdyugin\altaffilmark{2}, 
D. M. Fluri\altaffilmark{3}, 
V. Piirola\altaffilmark{2} }

\altaffiltext{1}{Kiepenheuer Institut f\"ur Sonnenphysik, Sch\"oneckstrasse 6, D-79104 Freiburg, Germany; sveta@kis.uni-freiburg.de }

\altaffiltext{2}{Tuorla Observatory, Department of Physics and Astronomy, University of Turku, V\"ais\"al\"antie 20, FIN-21500, Piikki\"o, Finland; andber@utu.fi, piirola@utu.fi }

\altaffiltext{3}{Institute of Astronomy, ETH Zurich, CH-8093 Zurich, Switzerland; fluri@astro.phys.ethz.ch}

\begin{abstract}
We report first multicolor polarimetric measurements ($UBV$ bands) for the hot Jupiters HD189733b and confirm our previously reported detection of polarization in the $B$ band (Berdyugina et al.\ 2008). The wavelength dependence of polarization indicates the dominance of Rayleigh scattering with a peak in the blue $B$ and $U$ bands of $\sim$10$^{-4}\pm10^{-5}$ and at least a factor of two lower signal in the $V$ band. The Rayleigh-like wavelength dependence, detected also in the transmitted light during transits, implies a rapid decrease of the polarization signal toward longer wavelengths. Therefore, the { nondetection by Wiktorowicz (2009), based on a measurement} integrated within a broad passband covering the $V$ band and partly $B$ and $R$ bands, { is inconclusive} and consistent with our detection in $B$. We discuss possible sources of the polarization and demonstrate that effects of incomplete cancellation of stellar limb polarization due to starspots or tidal perturbations are negligible as compared to scattering polarization in the planetary atmosphere. 
We compare the observations with a Rayleigh-Lambert model and determine effective radii and geometrical albedos for different wavelengths. { We find a close similarity of the wavelength dependent geometrical albedo with that of the Neptune atmosphere, which is known to be strongly influenced by Rayleigh and Raman scattering.} Our result establishes polarimetry as a reliable means for directly studying exoplanetary atmospheres.
\end{abstract}

\keywords{planetary systems --- polarization --- stars: individual (HD189733)}

\section{Introduction}

Direct detection of exoplanets with polarimetry opens the prospect for probing their atmospheres. A reflecting planet breaks the symmetry of the stellar radiation and conspicuously marks its presence in polarized light \citep[e.g.,][]{fb10,berd11}. Thanks to the differential nature of polarimetry, its dynamic range exceeds that of any other technique, but it is still a challenge to detect low signals.  

It is known that polarization properties of the light are generally wavelength  dependent \citep[e.g.,][]{stenflo05}, with Thomson scattering being one exception. Rayleigh scattering being an important opacity source in upper layers of cool atmospheres results in polarization strongly increasing toward the blue with a $\lambda^{-4}$ law. Combined with an angular dependence of polarization, this dictates an optimal range of wavelengths and scattering angles at which polarization can be successfully detected. 

There have been a few attempts to detect polarized reflected light from an exoplanet. \citet{hou06} and \citet[][hereafter L09]{luc09} observed hot Jupiters in $\tau$\,Boo and 55\,Cnc in a red filter of 590\,nm to 1000\,nm (maximum at 800\,nm). Due to relatively low statistics and incomplete orbital phase coverage, only standard deviations of Stokes $q$ and $u$ could be deduced: 5.1$\cdot$10$^{-6}$ and 2.2$\cdot$10$^{-6}$ for the two systems, respectively. \citet[][hereafter B08]{berd08} observed another hot Jupiter, HD189733b, which is twice as close to the parent star as, e.g., $\tau$\,Boo\,b. The Stokes parameters were measured in the Johnson $B$ band (370--550\,nm, maximum at 430\,nm). Despite a relatively low accuracy of individual measurements, a very high statistics of the data (about 100 nightly measurements) and an even distribution over orbital phases have enabled the detection of variable, phase-locked polarization of maximum  $\sim$2$\cdot$10$^{-4}$ and best-fit amplitudes in Stokes $q$ and $u$ of (1.5$\pm$0.3)$\cdot$10$^{-4}$ and (1.1$\pm$0.4)$\cdot$10$^{-4}$, respectively, for binned data. \citet[][hereafter W09]{wik09} also observed HD189733 but in the 400--675\,nm band (centered at 550\,nm). An upper limit of 7.9$\cdot$10$^{-5}$ in the polarization degree (5$\cdot$10$^{-5}$ in Stokes $q$) from only six nightly measurements was reported. At first glance, these measurements contradict to each other. However, as they were taken at different wavelengths, { the conclusion made from} their direct comparison was incorrect. Furthermore, only one of the six W09 measurements (which provided the upper limit) is at a phase near elongation where a significant polarization signal is expected based on our B08 ephemeris. Therefore, the { nondetection} of W09 is { inconclusive} and consistent with our detection.

In this paper, we show that the detected amplitudes and upper limits across the optical spectrum indicate the dominance of Rayleigh scattering in the atmospheres of hot Jupiters. This is also favored by transit spectroscopy in the optical by \citet[][hereafter, P08]{pont08} and \citet{lec08a,lec08b} and in the near infrared by \citet{sing09}. Here we report first multicolor polarimetric measurements in the $UBV$ bands for HD189733b. We confirm within the standard deviation the previously reported first detection of polarization in the $B$ band by B08 and determine reflecting properties of the planet. We also find the consistency between the data of W09 and our measurements in the $V$ band.

\section{Observations}\label{sec:obs}

The new observations were carried out in 2008 April 18--24 and August 2--9 with the  TurPol $UBVRI$ polarimeter mounted in the Cassegrain focus at the 2.5m Nordic Optical Telescope (NOT). The polarimeter was designed by \citet{piirola73,piirola88}. It consists of a half- (or quarter-) wave plate as retarder, calcite block, diaphragms, chopper, filter wheel, dichroic mirrors, and five photomultipliers as detectors for the $UBVRI$ bands. The calcite block splits the light onto parallel ordinary and extraordinary beams. Thus, the sky contributes to both beams and its polarization (e.g., due to dust in the Earth atmosphere or moonlight) can be exactly compensated.
This is a great advantage of TurPol to the polarimeters PlanetPol of \citet{hou06} and its copy POLISH of W09 which have to monitor very accurately a variable sky polarization.
The rapidly rotating chopper (25\,Hz) alternates integration of the ordinary and extraordinary beams at the same detector for each band simultaneously, which strongly diminishes systematic effects due to instrumental sensitivity, seeing and variable sky transparency. To measure linear polarization, the retarder was rotated at 22.5\degr\ intervals. Each pair of the observed Stokes $q$ and $u$ was calculated from eight exposures at different orientations of the retarder, which allowed us to avoid systematic errors due to imperfections of the retarder. 
Our definition of the $qu$-system is in accordance with the common agreement that positive and negative $q$ are in the north-south and east-west directions, respectively, while positive and negative $u$ are at an angle of 45\degr\ counter-clockwise from the positive and negative $q$.

To determine the instrumental polarization, we observed 26 nearby, bright, nonvariable, and nonpeculiar stars ($V$=4$^{\rm m}$--6$^{\rm m}$, $d<50$\,pc) of spectral classes A--G, which were expected to have no detectable intrinsic and interstellar polarization. Note that W09 observed only one standard star which resulted in a poor absolute calibration of his data. These stars were repeatedly observed at various parallactic angles to determine both the telescope and instrument polarization, the former  rotating during observations due to the alt-azimuthal mount and the latter being constant. Some stars were observed a few times on different nights to ensure the stability. In addition, for calibration of the polarization angle zero point, we observed highly polarized standard stars HD132052, HD161056, and HD204827. Total integration time of 1--2 hours for each star allowed us to measure instrumental polarization with the accuracy of (1--2)$\cdot$10$^{-5}$.  
Measurements for those standard stars which were used for the calibration are presented in Table~\ref{tab:stand}.

\begin{deluxetable}{lrrr}
\tabletypesize{\scriptsize}
\tablewidth{0pt}
\tablecaption{Polarization of standard stars. 
\label{tab:stand}}
\tablehead{
\colhead{Target}  & \colhead{$[q, u]_U\pm\sigma_U$} 
                  & \colhead{$[q, u]_B\pm\sigma_B$}
                  & \colhead{$[q, u]_V\pm\sigma_V$}
}
\startdata
HD67228   &  [3, 15] $\pm$ 13  &   [3, -12] $\pm$  9   &  [-12,   9] $\pm$   15  \\
HD85376   &  [0,  4] $\pm$ 4   &  [-5,   0] $\pm$  3   &   [-8,  -5] $\pm$   5   \\
HD110897  &  [3,  5] $\pm$ 12  &   [0,   5] $\pm$  5   &    [1,   8] $\pm$   7   \\
HD121560  &  [7,  7] $\pm$ 8   &   [4,  -4] $\pm$  4   &   [-2,  16] $\pm$   6   \\
HD126053  &          ...       &   [6,   3] $\pm$  7   &    [1,   1] $\pm$   10  \\
HD139641  &  [5, -8] $\pm$ 6   &   [0,  -1] $\pm$  4   &   [-2,   3] $\pm$   5   \\
HD150997  & [-1,  8] $\pm$ 9   & [-13,   3] $\pm$  6   &   [-4, -10] $\pm$   7   \\
HD174160  &          ...       &  [11,  -1] $\pm$  6   &  [-25,  19] $\pm$   9   \\
HD185395  & [-6,  4] $\pm$ 5   &  [-6,   8] $\pm$  4   &  [-10,  14] $\pm$   8   \\
\hline
HD121560  &[14, -13] $\pm$ 10  &  [-5,  -9] $\pm$  4   &   [-9,  13] $\pm$   6   \\
HD126053  &          ...       &  [14,   0] $\pm$  5   &   [-5,  -8] $\pm$   5   \\
HD157466  & [0,  -7] $\pm$  6  &   [4,  -5] $\pm$  4   &   [-5,   4] $\pm$   5   \\
HD174160  & [3,   0] $\pm$  4  &  [-6,  -5] $\pm$  4   &   [-9,  16] $\pm$   5   \\
HD177082  & [6,  -2] $\pm$  5  &   [0,  -1] $\pm$  4   &    [0,   4] $\pm$   6   \\
HD182807  &[-4,  -2] $\pm$  3  &  [-4,  -4] $\pm$  3   &   [-2,   6] $\pm$   4   \\
HD197076  &[-3, -15] $\pm$  3  &  [-7,  -6] $\pm$  3   &   [-2,  -3] $\pm$   3   \\
HD198390  &[-2,   1] $\pm$  4  &  [-5,  -4] $\pm$  4   &   [-7,  -6] $\pm$   4   \\
HD206826  & [2,  -5] $\pm$  6  &   [0,  -6] $\pm$  4   &   [-8,  -2] $\pm$   5   \\
HD212395  &[-10,-16] $\pm$  8  &  [-5,  -4] $\pm$  3   &  [-16,  -1] $\pm$   3   \\
HD216756  &[10,   3] $\pm$  4  &  [10,  -1] $\pm$  5   &    [9,   2] $\pm$   4   \\
HD218261  & [1,  -1] $\pm$  3  &  [-1,  -2] $\pm$  3   &  [-10,   4] $\pm$   4   \\
HD225239  & [1,   2] $\pm$  4  &  [-2,  -6] $\pm$  3   &    [1,   7] $\pm$   4   \\
\enddata
\tablecomments{Polarization is in units of 10$^{-5}$. The upper and lower parts are for April and August, 2008, respectively.}
\end{deluxetable}

The instrumental polarization deduced from these observations and expressed in polar coordinates (the polarization degree $P_{\rm ins}$ and the angle $\theta_{\rm ins}$) is as follows:
in April 2008\\
  $P_{\rm ins}(U)$ = (23.1$\pm$2.7)$\cdot$10$^{-5}$, $\theta_{\rm ins}(U)$ = 106.5 $\pm$ 3.3,\\
  $P_{\rm ins}(B)$ = (13.4$\pm$1.6)$\cdot$10$^{-5}$, $\theta_{\rm ins}(B)$ = 117.9 $\pm$ 3.4,\\
  $P_{\rm ins}(V)$ = (11.8$\pm$2.5)$\cdot$10$^{-5}$, $\theta_{\rm ins}(V)$ = 116.5 $\pm$ 5.9;\\
and in August 2008\\
  $P_{\rm ins}(U)$ = (24.2$\pm$1.4)$\cdot$10$^{-5}$, $\theta_{\rm ins}(U)$ = 110.2 $\pm$ 1.6,\\
  $P_{\rm ins}(B)$ = (13.8$\pm$1.1)$\cdot$10$^{-5}$, $\theta_{\rm ins}(B)$ = 116.2 $\pm$ 2.2,\\
  $P_{\rm ins}(V)$ = (11.9$\pm$1.3)$\cdot$10$^{-5}$, $\theta_{\rm ins}(V)$ = 121.1 $\pm$ 3.1.\\
We found a remarkable agreement between measurements carried out in April and August. This proves the stability of the total instrumental polarization during our observing runs as well as the robustness of our choice of standard stars. Note also that during our August run there was some presence of Sahara dust in the atmosphere. L09 reported that Sahara dust can produce sky polarization of order 10$^{-5}$ at large zenith distances. This is at the noise level of our measurements. Also, as explained above, the construction of the TurPol enables exact compensation of any background polarization which is not variable within the time of alternation between the ordinary and extraordinary beams (1/25\,s in our case). The very good agreement between the measurements of the instrumental polarization in April and August proves that this technique works very well. Moreover, since we increased total integration time in August due to increased opacity of the sky, the accuracy achieved in August was on average better than that in April. Therefore, we conclude that Sahara dust has not changed the polarization noise or introduced bias into our observations. 

\begin{figure}
\centering
\resizebox{8.5cm}{!}{\includegraphics{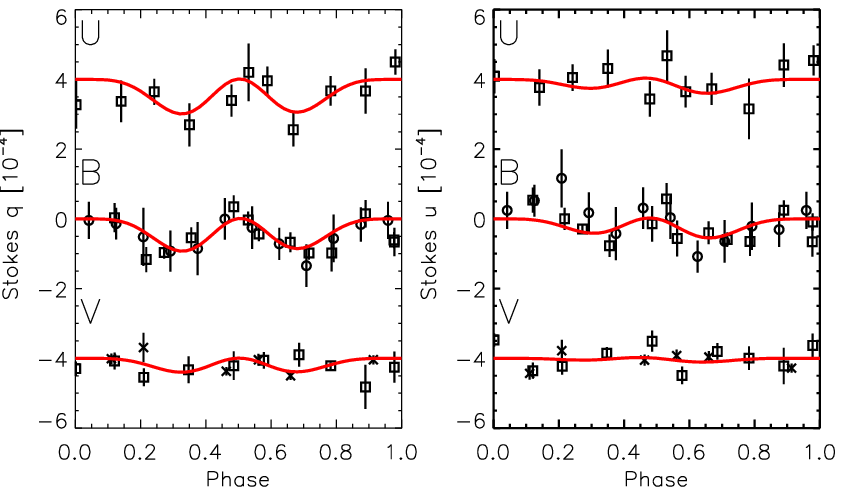}}
\resizebox{8.5cm}{!}{\includegraphics{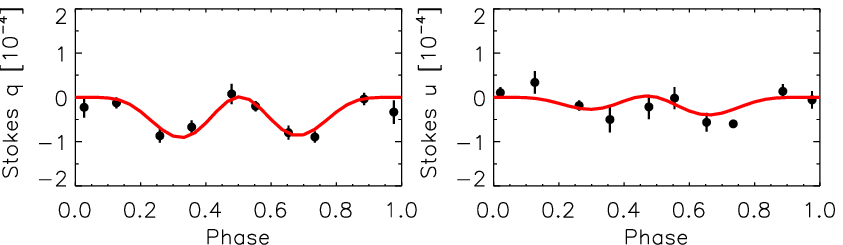}}
\caption{Stokes $q$ and $u$ with $\pm1\sigma$ error bars for HD189733b. 
{\it Upper panels}: the new $UBV$ measurements  are shown by {\it squares} and the binned $B$-band data from B08  by {\it open circles}, and the W09 measurements  by {\it crosses}. The $U$ and $V$ data are shifted in vertical by $\pm$4$\cdot$10$^{-4}$ for clarity.
{\it Lower panels}: All the $U$ and $B$ data from the years 2006--2008 binned together. The mean error of the binned data is 1.7$\cdot$10$^{-5}$, and the standard deviation is 1.1$\cdot$10$^{-5}$. 
Curves are the best-fit solutions for the Rayleigh-Lambert atmosphere. 
The normalized $\chi^2$ of the fit is 1.16.
}
\label{fig:pol}
\end{figure}

Our science targets were a number of nearby systems with hot Jupiters. Here we report results for HD189733b only, while data for other systems will be analyzed in subsequent papers. We made 10\,s exposures in the $UBV$ passbands simultaneously during 3--4 hours every clear night and obtained 1$-$2 measurements per night in each band (35 in total) with individual errors of (4--6)$\cdot$10$^{-5}$ in the $U$ band and (2--4)$\cdot$10$^{-5}$ in the $B$ and $V$ bands. The data were corrected for instrumental polarization. The calibrated measurements for the three passbands are shown in Fig.~\ref{fig:pol} (upper panels). The following amplitudes in the $UBV$ bands are detected (in the 10$^{-5}$ scale): 
in Stokes $q$, respectively, 
9.4$\pm$2.1,  
9.1$\pm$2.4, 
4.6$\pm$2.4; 
and in Stokes $u$
4.4$\pm$2.8, 
4.3$\pm$2.4, 
2.1$\pm$1.4.
The uncertainties were calculated as standard deviations from the best fits. The $B$ amplitude is in agreement with the previously reported polarization by B08 within the standard deviation. In Stokes $q$ the largest difference  is seen only in one B08 binned data point near the phase 0.7, while all other points agree well with the new data. In Stokes $u$ only two points deviate near the phases 0.2 and 0.65, but they are still within about one standard deviation. As the $U$ and $B$ amplitudes are similar, we can improve the statistics by binning together the new $UB$-band data and the $B$-band data of B08 (lower panels of Fig.~\ref{fig:pol}). The standard deviation is then reduced down to 1.1$\cdot$10$^{-5}$.

We also find a consistency between our measurements in the $V$ band and those of W09, which were centered at about the same wavelength  (Fig.~\ref{fig:pol}).
One can see that only one of the six W09 measurements (which provided the upper limit) is at a phase near elongation where a polarization signal of about 5$\cdot$10$^{-5}$ is expected. It matches perfectly the W09 upper limit. Therefore, the ``null'' result concluded by W09 is misleading due to the overlooked wavelength dependence of polarization.

\section{Origin of polarization}\label{sec:origpol}

Our primary hypothesis is that the observed polarization is due to scattering in the planetary atmosphere. The shape of the phase curve with two peaks near elongations is a strong argument in favor of this (Fig.~\ref{fig:pol}). It can be reproduced with a simple reflecting model (Sect.~\ref{sec:mod}). 

However, one can imagine alternative scenarios for the origin of the observed polarization. For example, starspots detected on the host star \citep[e.g.,][hereafter P07]{pont07} when appear near the limb may cause incomplete cancellation of the stellar limb polarization. Assuming typical parameters of starspots on a K dwarf \citep{berd05} and computing its most probable limb polarization at 400\,nm \citep{fluri_stenflo99}, we evaluated a possible effect for the large spot Feature A detected by P07.  Such a spot near the limb would cause maximum 5$\cdot$10$^{-7}$ polarization. If its area is artificially increased up to 1\%, which is the maximum spot area observed on the stellar disk \citep{winn07}, the polarization can be 3$\cdot$10$^{-6}$. Note that the projected spot area is strongly reduced toward the limb, and cannot be as large as 1\% near the limb (this is in contrast, e.g., to transits, when the projected area of the planet does not depend on the limb distance). Also, due to possible simultaneous presence of several spots near different parts of the limb, an expected effect is even smaller. These calculations demonstrate that the spot area on HD189733 is simply too small to be responsible for the 10$^{-4}$ effect in linear polarization. A transversal Zeeman effect due to starspot magnetic fields is also excluded at the level above 10$^{-6}$ \citep{moutou07}.

Another symmetry breaking effect resulting in a non-zero polarization could be due to tidal interaction of the planet with the star. Following \citet{condon_schmidt75} we have evaluated the height of tidal bumps on the stellar surface to be about 1\,km. This results in the ellipticity of the star of 2$\cdot$10$^{-6}$ and the maximum residual limb polarization of 10$^{-9}$, which is totally negligible. 

We therefore concentrate our efforts on modeling the observed polarization under the assumption that it is caused by scattering in the planetary atmosphere.

\begin{deluxetable}{lcccc}
\tabletypesize{\scriptsize}
\tablewidth{0pt}
\tablecaption{Parameters of the RL-atmosphere. 
\label{tab:res}}
\tablehead{
\colhead{} & \colhead{$U$} & \colhead{$B$} & \colhead{$V$} & \colhead{$RI$} 
}
\startdata
$\Omega$ [\degr]          &  10$\pm$15   &  14$\pm$6    &   5$\pm$20   &   -     \\ 
$R_{\rm RL}$/$R_{\rm J}$  &1.19$\pm$0.24 &1.18$\pm$0.10 &0.75$\pm$0.20 &   $<$0.43  \\
$p_{\rm RL}$         &$\le$0.62$\pm$0.30 &0.61$\pm$0.12 &0.28$\pm$0.16 &   $<$0.09  \\
\enddata
\tablecomments{$\Omega$ values differing by 180\degr are equally good. If our $V$ band data are combined with those of W09, $\Omega$=0\degr$\pm$9\degr, $R_{\rm RL}$=0.79$\pm$0.13, and $p$=0.31. 
}
\end{deluxetable}

\section{Rayleigh scattering in the planetary atmosphere}\label{sec:mod}

To model polarization of the light scattered off a planet we employ the Rayleigh-Lambert (RL) approximation, i.e.\ assuming (i) Rayleigh scattering for polarization and (ii) the Lambert sphere with the geometrical albedo $p_{\rm L}=$2/3 for intensity \citep[][B08]{seager00,fb10}. Since the albedo of the atmosphere is fixed, the wavelength dependent properties of the planet are effectively included into the radius of the RL-atmosphere $R_{\rm RL}$ for different bands which represents a geometrical limit for the unity optical thickness in the atmosphere. Note that this model implies a Bond albedo of 1 and scattered light maximally polarized, so the values for radii and albedos should only be considered as limits for a spherically symmetric case. We use the same input parameters and the $\chi^2$ minimization procedure as in B08 and \citep{fb10} and apply it to the original measurements obtained in 2006--2008 for each passband separately. For the orbit inclination $i$ we used 94.32\degr\ as it results in a smaller $\chi^2$ value as compared to the complementary value of 85.68\degr\ reported by P07. The best-fit values of the two free parameters of the model $\Omega$ and $R_{\rm RL}$ are listed in Table~\ref{tab:res}. The modeled Stokes $q$ and $u$ curves are shown in Fig.~\ref{fig:pol}. The values for the $B$ band agree with those obtained by B08 within the standard deviation. If our $V$-band data are combined with those of W09, the parameters are about the same but the errors improve.

\begin{figure}
\centering
\resizebox{8.5cm}{!}{\includegraphics{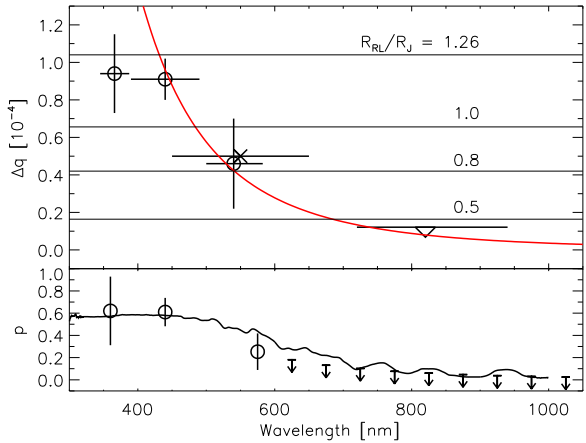}}
\caption{Polarimetric amplitudes { $\Delta q$ ({\it top}) and geometrical albedo $p$ ({\it bottom}) measured} in various passbands. 
Our new data are shown by {\it open circles}, the W09 upper limit by a {\it cross}, and the L09 upper limit for $\tau$~Boo~b scaled to the orbit of HD189733b by a {\it triangle}. Horizontal bars show the passbands' FWHI. 
{
{\it Upper panel:}
Thin horizontal lines indicate $\Delta q$ for $R_{\rm RL}$/$R_{\rm J}$ = 1.26, 1.0, 0.8, 0.5 and $\Omega$=14\degr\ \citep{fb10}; {\it solid} (red) curve is the Rayleigh law $\Delta q = 3.6\cdot10^6\lambda^{-4}$ scaled to fit the $BVRI$ data.
{\it Lower panel:} 
{\it Solid} curve is the geometrical albedo of Neptune smoothed by a 100\,nm boxcar.
{\it Downward arrows} are upper limits obtained from the solid curve of the upper panel and the transit data by P08.
}
Our data are consistent with other measurements and reveal Rayleigh scattering in the planetary atmosphere. 
}
\label{fig:amp}
\end{figure}

To evaluate a wavelength-dependent geometrical albedo (Table~\ref{tab:res}), we compare $R_{\rm RL}$ with the results by P08. They employed analytical transit curves by \citet{mandelagol02} under the assumption that the planetary body is fully opaque which provides lower limits at observed wavelengths. They modeled the transit curves by allowing the equivalent radius of the planet, $R_{\rm eq}$, to vary with wavelength and found an increase of the radius by $\sim$500\,km at 0.55\,$\mu$m as compared with 1\,$\mu$m, and by $\sim$1000\,km as compared with mid-infrared. Using $R_{\rm eq}$=1.1524\,$R_{\rm J}$ for 0.55\,$\mu$m ($R_*$=0.755$R_\odot$),
we evaluate the geometrical $V$-band albedo of 0.28 by scaling $p_V=p_{\rm L}R_{\rm RL}^2(V)/R_{\rm eq}^2(V)$. 
The transit in the $B$ band (0.44\,$\mu$m,) was observed about 0.4\% deeper than in $V$ \citep{bou05}, which resulted in $R_{\rm eq}=1.26 R_{\rm J}$, but it was deduced with different orbital and stellar parameters than used by P08. By applying the P08 parameters, we obtain $R_{\rm eq}(B)=1.23 R_{\rm J}$. This leads to $p_B$=0.61, which is a factor of two larger than that for the $V$-band. In the $U$ band $R_{\rm eq}$ is unknown but can be significantly larger than in the optical, as found for other hot Jupiters in the UV \citep[e.g.,][]{vm03,lin10,lec10}. Using $R_{\rm eq}(B)$, an upper limit of $p_U$ is 0.62. This high albedo in the blue is consistent with model predictions for hot Jupiters \citep{bur08}. Using $R_{\rm eq}(V)$, the upper limit for both $p_U$ and $p_B$ is 0.7. For red bands ($RI$) we can use the upper limit of the polarization amplitude obtained by L09 for $\tau$~Boo~b but scaled to the orbit of HD189733b, which results in 1.2$\cdot$10$^{-5}$ (Fig.~\ref{fig:amp}) and $R_{\rm RL}=0.43 R_{\rm J}$. Using $R_{\rm eq}=1.1469 R_{\rm J}$ for 0.85\,$\mu$m from P08, $p_{RI}$ is 0.09, in agreement with upper limits for hot Jupiters in optical red wavelengths \citep[e.g., L09,][]{rowe08}. Note that the absolute values of $R_{\rm eq}$ and $p$ rely on $R_*$=0.755$R_\odot$, while a direct measurement of the stellar angular diameter resulted in 0.779$\pm$0.052$R_\odot$ \citep{baines2007}. Using this value would increase radii and decrease albedos. Also, the geometrical albedo derived from polarimetry does not include unpolarized thermal emission of the planet which starts to dominate at longer wavelengths \citep[e.g.,][]{bur08}.

The wavelength dependence of $R_{\rm RL}$ and $p$ indicates that optical polarization is dominated by Rayleigh scattering with a characteristic increase of the cross section toward the blue. The same conclusion was reached by \citet{lec08a} who modeled the P08 transit data with a translucent atmosphere (single layer). They convincingly showed that the main optical opacity source in the HD189733b atmosphere must have a cross section with the $\lambda^{-4}$ power law and suggested two possible scattering sources, H$_2$ molecules and MgSiO$_3$ condensate. The the radius change from the red to the blue of such an translucent atmosphere increased by a factor of two, up to about 1000\,km, as compared to the fully opaque case of P08. Note that extrapolation of the planet radius using this model to the near infrared fits well the corresponding transit measurements \citep{sing09}, but it underestimates the radius in the $B$ band (and probably in $U$). This can be due to additional opacity sources in the optical, e.g., H$^-$ absorption which increases at 0.3--0.9\,$\mu$m and decreases at 0.9--1.6\,$\mu$m.

{ The $\lambda^{-4}$ law fits well the $BVRI$ polarization amplitudes and upper limits (Fig.~\ref{fig:amp}).} Thus, it appears that { most optical photons ($\lambda$$>$400\,nm) are scattered only once in the atmosphere of HD189733b, because} in such a case the polarization amplitude is proportional to the scattering cross section \citep[e.g.,][]{stenflo05}. Therefore, models with multiple scattering of all photons cannot explain the observed polarization in the blue as they predict the maximum polarization for HD189733b of  (3--4)$\cdot10^{-5}$  \citep[e.g.,][L09; using the $V$-band transit radius]{seager00,stam04}. 

The $U$ band polarization amplitude deviates from the Rayleigh law.
{ 
Interestingly, even though the effective temperature of the Solar system planets are cooler than this planet, we find that $p(\lambda)$ for Neptune \citep{karkoschka1994} is quite similar to HD189733b (Fig.~\ref{fig:amp}, lower panel), which is also similar to that of Uranus but somewhat different from Jupiter and Saturn. The reflected and polarized spectra of Neptune are well known to be strongly influenced by Rayleigh and Raman scattering by H$_2$ as well as methane absorption in the red \citep[e.g.,][]{sromovsky2005,joos2007}. Rayleigh scattering results in the characteristic decrease of the albedo and polarization in the continuum  at $\lambda$$>$500\,nm, while Raman scattering reduces the reflectivity of the atmosphere at shorter wavelengths due to partial absorption of the energy of blue photons. Absorption by molecules is also possible. Thus, enhanced blue absorption can explain the $U$-band polarization and albedo of HD189733b, while Raman scattering can be an additional opacity source leading to heating of exoplanetary atmospheres. 

The atmosphere of Neptune can be approximated by a high-altitude haze layer above a semiinfinite cloud deck \citep{sromovsky2005}. Similarly, a preliminary semi-empirical model atmosphere of HD189733b \citep{berd11} explains the observed polarization by the presence of a distinct dust condensate layer unbeneath a thin gaseous layer. The details of this model will be presented in our forthcoming paper.
}


\section{Conclusions}\label{sec:con}

The new $UBV$ polarimetric observations of HD189733b confirm the first detection of polarized reflected light from the hot Jupiter by B08 at the 10$^{-4}$ level with the standard deviation of 10$^{-5}$. Our data are consistent with upper limits obtained by others at different wavelengths. They clearly demonstrate the dominance of Rayleigh scattering in the planetary atmosphere at optical wavelengths longer than 400\,nm.
At shorter wavelengths an additional opacity mechanism { (e.g., Raman scattering)} plays a significant role. Our result establishes polarimetry as an important tool for studying directly exoplanetary atmospheres in the visible and near UV. In the near future it will be employed for non-transiting systems.

We appreciate an extended review by an anonymous referee.
SVB acknowledges the EURYI (European Young Investigator) Award provided 
by the ESF (see www.esf.org/euryi) and the SNF grant PE002-104552. This work is also supported by the Academy of Finland, grant 115417.
Based on observations made with the Nordic Optical Telescope, La Palma, Spain.

\end{document}